\documentstyle[times,epsfig]{mn}
\input{epsf}

\newcommand{\bez}{\begin{eqnarray*}}
\newcommand{\eez}{\end{eqnarray*}}
\newcommand{\be}{\begin{equation}}
\newcommand{\ee}{\end{equation}}
\newcommand{\beq}{\begin{eqnarray}}
\newcommand{\eeq}{\end{eqnarray}}
\newcommand{\bc}{\begin{center}}
\newcommand{\ec}{\end{center}}

\newbox\grsign \setbox\grsign=\hbox{$>$} \newdimen\grdimen \grdimen=\ht\grsign
\newbox\simlessbox \newbox\simgreatbox \newbox\simpropbox
\setbox\simgreatbox=\hbox{\raise.5ex\hbox{$>$}\llap
     {\lower.5ex\hbox{$\sim$}}}\ht1=\grdimen\dp1=0pt
\setbox\simlessbox=\hbox{\raise.5ex\hbox{$<$}\llap
     {\lower.5ex\hbox{$\sim$}}}\ht2=\grdimen\dp2=0pt
\setbox\simpropbox=\hbox{\raise.5ex\hbox{$\propto$}\llap
     {\lower.5ex\hbox{$\sim$}}}\ht2=\grdimen\dp2=0pt

\begin{document}
{

\title[Statistical constraints on non-cosmological subclasses of GRBs]
{Statistical constraints on non-cosmological subclasses of GRBs}

\author[Ya.~Tikhomirova, B.~E.~Stern and R.~Svensson]
{\parbox[]{6.8in} {Yana~Tikhomirova,$^{1,2\star}$ 
Boris~E.~Stern$^{3,1,2\star}$
and Roland~Svensson$^{2\star}$}\\
$^1$Astro Space Centre of Lebedev Physical
Institute, 84/32 Profsojuznaja Street, Moscow 117997, Russia\\
$^2$SCFAB, Stockholm Observatory, Department of Astronomy,
SE-106 91 Stockholm, Sweden \\
$^3$Institute for Nuclear Research of Russian Academy of Sciences,
7a, Prospect 60-letija Oktjabrja, Moscow 117312, Russia}

\date{Accepted, Received}

\maketitle


\begin{abstract}
\noindent
There still exists the possibility that the phenomenon of 
gamma-ray bursts (GRBs) is a mixture of events of different nature, even  
within the class of long ($>$2 s) bursts.
We try to put statistical constraints on 
a possible non-cosmological component  
using the uniform GRB catalog of Stern \&Tikhomirova$^\#$
obtained from an overall scan of the full 9.1 year BATSE 1024 ms data. 
The sample consists of 3906 GRBs and includes 
non-triggered bursts with peak fluxes down to 0.1 photons cm$^{-2}$ s$^{-1}$.
We find no significant deviations from isotropy. The constraints 
on a non-cosmological population are still weak. The allowed contribution of 
a GRB subpopulation originating from an extended galactic halo is $\sim$ 60\% 
and the upper limit on an Eucledian component (e.g., nearby galactic or 
non-cosmological extragalactic GRBs) is 23\%.  
The results concern mainly the class of long GRBs.

\end{abstract}


\section{Introduction}

 At present there are direct redshift measurements for 23 
Gamma-Ray Bursts$^\&$).
All of them are cosmological. If we believe that all GRBs
are of the same nature, then we must reject all
non-cosmological models of GRBs and to close this issue.

\footnotetext{$^\star$ E-mail: 
jana@anubis.asc.rssi.ru (YT); stern@lukash.asc.rssi.ru (BES); 
svensson@astro.su.se (RS)}

\footnotetext{$^\#$ available at
http://www.astro.su.se/groups/head/grb\_archive.html}
 
\footnotetext{$^\&$ see Greiner' web page
at http://www.aip.de/~jcg/grb.html}

However, first we should mention that long ($>$2 s) and
short ($<$2 s) GRBs are probably events of different nature
(Kouveliotou et al. 1993) and all of the redshift
measurements have been done only for long bursts.
Moreover, the impression of the uniformity of GRBs could
arise from their diversity whereas they can be
a mixture of events of different origin even within
the class of long bursts.
For example, Horvath (1998) suggested
the existance of an ``intermediate'' class of GRBs
with durations of several seconds and softer spectra.
The evidence for this is still not statistically
convincing. Nevertheless such a possibility should not be discarded.

In this work, we revisit statistical studies 
of non-cosmological models which 
were done before the discovery of afterglows. 
Now we have a much larger 
statistics which will not be extended during the next few years. 
Therefore it is interesting to outline constraints
on non-cosmological (i.e., local galactic, galactic halo,
or low redshift extragalactic) GRB subpopulations
which we can obtain from the present GRB sample. We concentrate
only on those models which can provide a reasonable isotropy of GRBs
and therefore we do not consider the galactic disk population 
(except for a very local one). 

Previous attempts to constrain Galactic GRB scenarios
(see, for example, Hakkila et al., 1994, Loredo \& Wasserman, 1998)
showed that the least constrained local model
is an extended galactic halo of bursters.
According to Hakkila et al. (1994), for some narrow range of parameters,
all GRBs could originate from a galactic halo being still in agreement
with the angular and the brightness distributions of GRBs
as observed by the Burst And Transient Source Experiment 
(BATSE)(Fishman 1992).

It is also interesting to recalculate a possible Euclidean component
(i.e., a population homogeneously distributed in Euclidean space).
It could be represented by a local Galactic population at distances
$<$500 pc or by extragalactic sources at z$\ll$1.
The estimate of Kommers et al. (2000) for the BATSE sample
with peak fluxes down to $\approx$ 0.2 photons cm$^{-2}$ s$^{-1}$
is about 10\%.

The deepest and largest GRB sample was found by search for
non-triggered GRBs in the BATSE continuous records of Stern et al.
(2000, 2001). Their new uniform catalog$^\#$ (UC)
includes 3906 (2068 triggered and 1838 non-triggered) events
with peak fluxes down to $\approx$ 0.1 photons cm$^{-2}$ s$^{-1}$
(a factor 2 lower than the BATSE trigger threshold).
Moreover, they applied a new method to measure the efficiency
of the search which is important for the reconstruction of
the real log $N$-log $P$ distribution.

Note, that all bursts of the UC were extracted from the DISCLA BATSE
data of the 1024 ms time resolution. The bursts which can be classified
as short (1 bin events) are about 12\% in the sample.
Therefore, {\bf the constraints
will concern mainly the class of long GRBs}.
In \S2, we present the results of a general test of the UC
for large scale isotropy. We constrain
the possible contribution of a halo population in \S3, and
of an Euclidean component in the GRB statistics in \S4.

\begin{figure}
\centerline{\epsfig{file=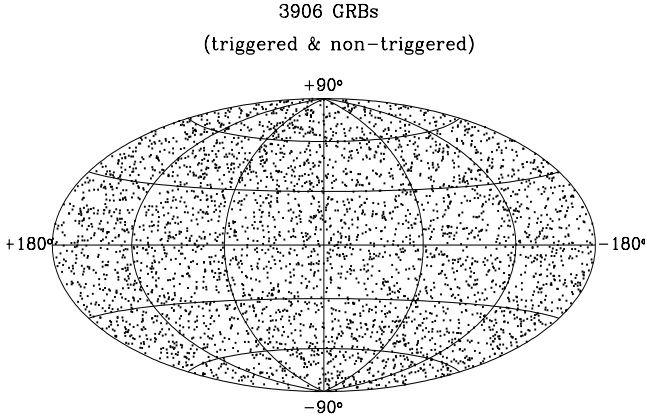,width=8.0cm,height=5.5cm}}
\caption{Sky distribution of the 3906 GRBs in the UC 
on an Aitoff-Hammer projection in Galactic coordinates.}
\end{figure}

\section{Test for isotropy}

The preliminary test of the UC for isotropy was done
in Tikhomirova \& Stern (2000). Here, we extend the analysis
to the complete sample.

\subsection{Statistical tests}

Figure 1 shows the distribution of the locations 
of all UC GRBs in galactic coordinates.
We use the following tests for the large scale isotropy 
(see Hartmann \& Epstein 1989 for test 1) and Briggs 1993, 1994
for tests 2)-5)) :\\
1) The dipole vector $\overrightarrow{R}$:
\begin{displaymath}
\overrightarrow{R}=\frac{\sum_i \overrightarrow{r_{i}}}{N},
\end{displaymath}
where $\overrightarrow{r_{i}}$ is the unit vector 
in the direction of a GRB and $N$ is the number of bursts 
in the sample. $\overrightarrow{R}$ characterizes the  dipole moment
in a coordinate free way and tests for 
large scale anisotropy.\\
2) $\langle\cos(\theta)\rangle$, where $\theta$ is the angle between
the Galactic center and a GRB. This quantity characterizes the dipole moment
in the Galactic system and tests for a Galactic population.\\
3) $\langle\sin^2(b)-1/3\rangle$, where $b$ is the Galactic latitute 
of a burst. This quantity characterizes the quadrupole moment in the
Galactic system and tests for a local Galactic population.\\
4) $\langle\sin(\delta)\rangle$, where $\delta$ is the declination of a burst.
and\\
5) $\langle\sin^2(\delta)-1/3\rangle$, which characterizes the dipole and quadrupole
moments in the equatorial system and tests for systematic errors.

The statistical errors are
$\sigma = (1/\sqrt{3N})$
for the components of $\overrightarrow{R}$ and
the tests 2) and 4); and
$\sigma = (\sqrt{4}/\sqrt{45N})$
for the tests 3) and 5). The statistical error
for the absolute value of $\overrightarrow{R}$ is
$\sigma = (1/\sqrt{N})$.
The total error has an additional component
associated with the errors of the locations of individual GRBs.
Although the location accuracy for the UC GRBs is
as low as several degrees (Tikhomirova \& Stern 2000) 
this component is smaller by an order of magnitude 
than the statistical one as was checked using Monte-Carlo simulations.

\subsection{Expected values}

The values of $|\overrightarrow{R}|$ and of the
statistics 2)-5) are zero for an isotropic distribution
of bursts observed by an ideal instrument. The values corrected
for the non-uniform BATSE sky exposure (Paciesas et al. 1999) 
are listed in the second column of Table 1. 
In the BATSE catalogs, these quantities usually have their expected
values (see Paciesas et al. 1999).
However, other systematic effects may appear 
as a result of a non-uniform burst selection.
One such effect is caused by an active CygX-1, which
reduces the efficiency of weak GRBs detected in that direction.

Such effects may be revealed by the test bursts used 
in the scan of Stern et al. (2000, 2001) to estimate 
the scan efficiency. Test bursts are initially
isotropic and are subjected to the same systematic 
effects as real bursts. However, their statistics is
limited ($\sim$ 5300). Therefore, we consider both the values
expected for the BATSE sky exposure (column 2 in Table 1)
and the values observed for test bursts (column 4 in Table 1). 

\subsection{Results}

\begin{table*}
\vskip 1.0cm
{Table 1. Results of the large scale isotropy test}.\\
\vskip 0.5cm
\begin{tabular}{ccccccc}
\hline
\hline
&expected&4Br&\multicolumn{4}{c}{UC}\\
\cline{4-7}
statistics&for&catalog&test bursts&all bursts&triggered&non-triggered\\
&isotropy&(1637)$^*$&(5345)$^*$&(3906)$^*$&(2068)$^*$&(1838)$^*$\\
\hline
\hline
$|\overrightarrow{R}|$
& 0.018
& 0.029
& 0.026
& 0.021
& 0.020
& 0.030 
\\
&&$\pm$0.025
&$\pm$0.014
&$\pm$0.016
&$\pm$0.022
&$\pm$0.023
\\
R1
& 0.000
& 0.005
&$-$0.022
&$-$0.012
&$-$0.006
&$-$0.019
\\
&&$\pm$0.014
&$\pm$0.008
&$\pm$0.009
&$\pm$0.013
&$\pm$0.013
\\
R2
& 0.000
& 0.015
& 0.014
& 0.014
& 0.005
& 0.024
\\
&&$\pm$0.014
&$\pm$0.008
&$\pm$0.009
&$\pm$0.013
&$\pm$0.013
\\
R3
& 0.018
& 0.024
& 0.003
& 0.010
& 0.019
& $-$0.000
\\
&&$\pm$0.014
&$\pm$0.008
&$\pm$0.009
&$\pm$0.013
&$\pm$0.013
\\
\hline
$\langle\cos(\theta)\rangle$
&$-$0.009
&$-$0.025
&$-$0.012
&$-$0.016
&$-$0.013
&$-$0.019
\\
&&$\pm$0.014
&$\pm$0.008
&$\pm$0.009
&$\pm$0.013
&$\pm$0.013
\\
\hline
$\langle\sin^2(b)-1/3\rangle$
&$-$0.004
&$-$0.001
&0.001
&$-$0.007
&$-$0.007
&$-$0.006
\\
&&$\pm$0.007
&$\pm$0.004
&$\pm$0.005
&$\pm$0.007
&$\pm$0.007
\\
\hline
$\langle\sin(\delta)\rangle$
& 0.018
& 0.024
& 0.003
& 0.010
& 0.019
& $-$0.000
\\
&&$\pm$0.014
&$\pm$0.008
&$\pm$0.009
&$\pm$0.013
&$\pm$0.013
\\
\hline
$\langle\sin^2(\delta)-1/3\rangle$
& 0.024
& 0.025
& 0.022
& 0.025
& 0.024
& 0.026
\\
&&$\pm$0.007
&$\pm$0.004
&$\pm$0.005
&$\pm$0.007
&$\pm$0.007
\\
\hline
\hline
\multicolumn{7}{l}{\sf $^*$ \footnotesize the number of events in each sample}
\end{tabular}
\end{table*}

The results of the tests together with the 1$\sigma$ statistical errors 
are listed in columns 4-7 of Table 1.  
The values for the 4Br catalog are given for comparison in
column 3. (Note, that $\overrightarrow{R}$ was not given in the 4Br catalog.)
Test bursts show a marginally significant dipole (at the $\sim 2.5\sigma$ level)
in the direction opposite to Cyg X-1, which is caused by a strong
deficit of detected weak bursts in a cone $\sim 25^{\rm o}$ around
Cyg X-1. A similar effect appears for the sample of real GRBs.

Table 2 shows the deviations from the values
expected for the BATSE exposure (columns 2-5) and from the
values for test bursts (columns 6-8). 
The deviations from the value for the test bursts
are given in units of $\sigma$ determined as
\begin{equation}
\sigma = \sqrt{\sigma_{\rm test~bursts}^2 + \sigma_{\rm real~bursts}^2}.
\end{equation}
All deviations are within $1.5\sigma$ so the results of the tests are
consistent with isotropy.

\begin{table*}
\vskip 1.0cm
{Table 2. Deviations from the expected values 
for large scale isotropy}.\\
\vskip 0.5cm
\begin{tabular}{cccccccc}
\hline
\hline
&\multicolumn{4}{c}{deviation from the value}
&\multicolumn{3}{c}{deviation from the value}\\
statistics&\multicolumn{4}{c|}{expected for the isotropy}
&\multicolumn{3}{c}{for the test bursts}\\
&4Br&\multicolumn{3}{c}{UC}
&\multicolumn{3}{c}{UC}\\
\cline{3-8}
&catalog&all&trig.&\multicolumn{1}{c}{non-tr.}&all&trig.&non-tr.\\
\hline
\hline
\rm
$\overrightarrow{R}$$^*$ 
&$0.7\sigma$&$1.2\sigma$&$0.3\sigma$&$1.5\sigma$
&$0.6\sigma$&$0.9\sigma$&$0.4\sigma$\\
\hline
$\langle\cos(\theta)\rangle$
&$-1.1\sigma^{**}$&$-0.7\sigma$&$-0.3\sigma$&$-0.8\sigma$
&$-0.3\sigma$&$-0.1\sigma$&$-0.5\sigma$\\
\hline
$\langle\sin^2(b)-1/3\rangle$
&$+0.4\sigma^{**}$&$-0.5\sigma$&$-0.5\sigma$&$-0.2\sigma$
&$-1.1\sigma$&$-1.0\sigma$&$-0.8\sigma$\\
\hline
$\langle\sin(\delta)\rangle$
&$+0.4\sigma^{**}$&$-0.9\sigma$&$+0.1\sigma$&$-1.3\sigma$
&$+0.6\sigma$&$+1.1\sigma$&$-0.2\sigma$\\
\hline
$\langle\sin^2(\delta)-1/3\rangle$
&$+0.1\sigma^{**}$&$+0.2\sigma$&$+0.0\sigma$&$+0.3\sigma$
&$+0.5\sigma$&$+0.3\sigma$&$+0.5\sigma$\\
\hline
\hline
\multicolumn{8}{l}{\sf $^*$ \footnotesize deviation of the vector}\\
\multicolumn{8}{l}{\sf $^{**}$ \footnotesize Paciesas et al. (1999)}
\end{tabular}
\end{table*}

\noindent

We also checked the excess of bursts towards M31 for the full
sample in the UC as well as for only weak 
($<$0.4 photons cm$^{-2}$ s$^{-1}$) bursts.
There is no excess.              

\section{Test for an extended galactic halo subpopulation of GRB sources}

\subsection{The Model for an Extended Halo of GRB sources}

The traditional model of a halo is associated with old neutron stars 
which have been ejected from the galactic disk 
(Shklovskii \& Mitrofanov 1985).
The test for an extended halo population is based on the displacement 
of the Solar system from the Galactic center (8.5 kpc) and 
the presence of the nearby galaxy M31 (670 kpc away) 
which should have a similar extended halo. 
Then we must check two things:
the dipole moment in the direction of the Galactic center and the excess of 
bursts around M31. The larger the halo of bursters, the smaller is 
the dipole moment but then GRBs from M31 become more visible. 
The same logic was used in previous works 
(e.g., Hakkila et al., 1994).
    
In the approximation of a steady outflow of neutron stars,
we have a spatial distribution of bursters with 
a Coulomb-like tail and a core:
\begin{equation}\label{spatdistrib}
N(r)=N_0 r_c^2/(r_c^2+r^2), 
\end{equation}
where $N(r)$ is the burster density per unit volume, 
$N_0$ is the density at the center, $r$ is the distance
from the Galactic center, and $r_c$ is the core size.
The distribution should decline faster than $1/r^2$ at some $r$
because a neutron star cannot emit GRBs for an infinite time or 
just because of the finite age of the Galaxy.
Expression (\ref{spatdistrib}) gives a
suitable asymptotic slope of the log $N$ - log $P$ distribution:
$dN/d(\log P) \propto P^{-\alpha}$ with $\alpha = 0.5$ 
for small brightnesses $P$ in agreement with the observed data
(Kommers et al. 2000, Stern et al. 2001).

We assumed identical halos for both our Galaxy and M31 and
tried two cutoffs for the distribution (\ref{spatdistrib}): at 300 kpc
(i.e., halfway between the Galaxy and M31) and at 800 kpc which could
represent the case of an infinite distribution 
(we are within the M31 halo in this case). 
This rough model seems to be sufficient for
an approximate estimate. 

The GRB luminosity function was assumed to be 
a lognormal distribution: 
\begin{equation}\label{lumin}
dN/dL = \exp[-\log^2(L/L_0)/\sigma_{\rm L}^2],  
\end{equation}
where $L$ is the GRB peak luminosity, $L_0$ is the average GRB 
peak luminosity, and $\sigma_{\rm L}$ is the width of the GRB 
luminosity function.

The model distributions of GRBs were obtained using 
Monte-Carlo simulations.
The variable parameters were: the core radius, $r_c$, 
the average GRB peak luminosity, $L_0$, and the width of the GRB luminosity 
function, $\sigma_{\rm L}$. 
The averaged peak luminosity $L_0$ was expressed 
as $P_{100}$ measured in units of 
$2 \cdot 10^{41}$ erg s$^{-1}$ in the 50 - 300 keV energy range.
This is the luminosity which produces a flux of 1 photon s$^{-1}$ cm$^{-2}$ 
at a distance of 100 kpc ($L_0\approx 2 \times 10^{41}P_{100}$).

 The spatial distribution of GRBs was sampled according 
to (\ref{spatdistrib}), where $r$ is the distance of a GRB 
from the center of the Galaxy or from the center of 
M31 (with equal probabilities). 
The locations of GRBs on the sky were sampled imposing experimental
location errors depending on the ''observed'' brightness.
The intrinsic peak brightness of a GRB was 
sampled according to equation (\ref{lumin}). 
The ``observed'' distribution of simulated GRBs
was folded with the BATSE exposure function (Paciesas et al., 1999) and 
with the detection efficiency 
determined using the test bursts method of Stern et al. (2000, 2001). 

\subsection{The constraints on the halo subpopulation}

We represented the full sample of GRBs as 
$(\xi-1)S_0 + \xi S_h$ where $S_0$ is an isotropic 
subsample (presumably cosmological) with 
an unknown log $N$ - log $P$ distribution, and
$S_h$ is the model halo subsample.
We want to determine the value of the halo fraction $\xi$ for which 
the following four criteria are satisfied: 
\begin{enumerate}
\item The dipole moment, $D_g$, towards the Galactic center should be within 
the observed $2\sigma$ upper limit: $C_g = 0.009$
\item The fraction of bursts with locations within 25$^{\rm o}$ 
from the M31 location, $N_{25}$, 
should not exceed the observed $2\sigma$ upper limit which is 
$C_{25} = (N_{\rm real}-N_{\rm exp}+2\sigma)/N_{\rm real} = 0.14 $
\item The same should apply for an 18$^{\rm o}$ region around M31,
for $N_{18}$: $C_{18} = 0.21$
\item The log $N$ - log $P$ distribution 
should be consistent with the observed one:\\
-- the sum of the distributions for subsamples $S_0$ and $S_h$
should give the observed one,\\
-- the distribution of the subsample $S_0$ should be smooth, 
i.e., the difference between 
neighbouring points should not exceed $3\sigma$, and \\
-- the distribution of $S_0$ should not bend down too sharply at low
brightnesses, i.e., not steeper than $dN/d(\log P) 
\propto P^{+0.5}$ (see Fig. 2). This limiting slope corresponds to
a reasonably sharp turnover of the log $N$-log $P$ distribution
at its dim end.
\end{enumerate}

For the observed distribution, we used the log $N$ - log $P$ 
curve corrected for
the  detection efficiency of the scan of Stern et al. (2001).
All limits were calculated for the full sample of the UC
(3906 GRBs with peak fluxes down to 0.1 photons cm$^{-2}$ s$^{-1}$).  

We searched for the fraction $\xi$ at which 
the full model sample of GRBs satisfies all four criteria.
For the first three criteria, $\xi$
is just the minimum of the ratios $C_g/D_g$,  
$C_{25}/N_{25}$, and $C_{18}/N_{18}$.

We also tried to apply criteria 2 and 3 for weak bursts only
(with peak fluxes $<$0.4 photons cm$^{-2}$ s$^{-1}$). This case does not give 
stronger constraints.

\subsection{Results}

\begin{figure}
\centerline{\epsfig{file=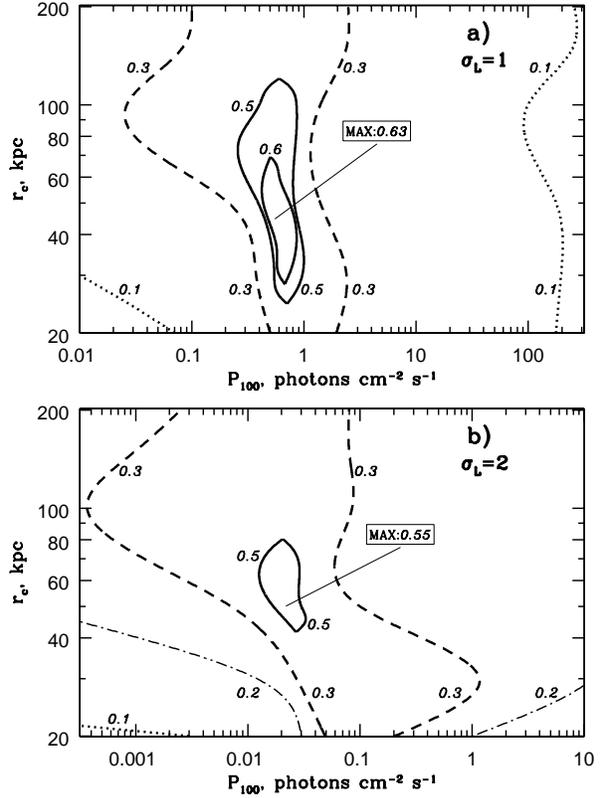,width=11cm,height=11cm}}
\caption{Maps of constant levels of the allowed fraction $\xi$
of the halo subpopulation in the whole sample of the UC:
a) for $\sigma_{\rm L}=1$, b) for $\sigma_{\rm L}=2$.
$r_c$ is the halo core radius, and $P_{100}$ is the GRB peak 
count rate at a distance of 100 kpc.}
\end{figure}

\begin{figure*}
\centerline{\epsfig{file=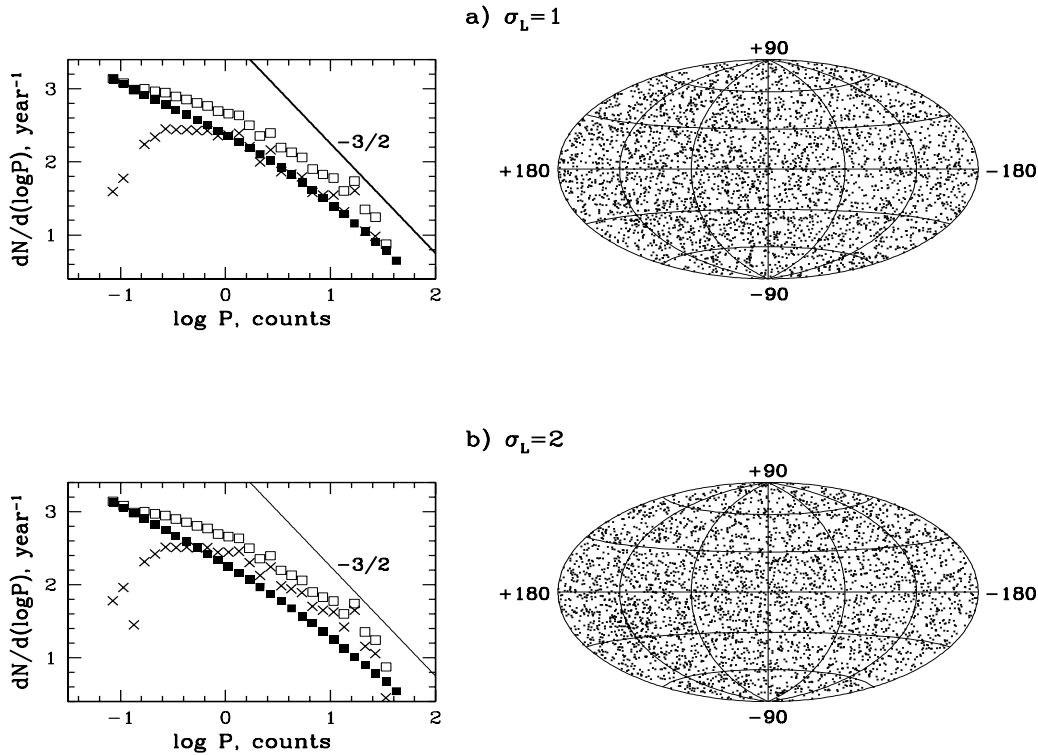,width=13cm,height=10cm}}
\caption{The simulated distributions  
for the case of maximum halo fraction $\xi$: 
a) for $\sigma_{\rm L}=1$, b) for $\sigma_{\rm L}=2$. 
Panels on the left, the log $N$ - log $P$ distributions: 
the empty squares represent the observed data according to the UC,
the filled squares represent the halo population, and 
the crosses correspond to the cosmological population.
Panel on the right, the sky distribution: the whole
sample (cosmological + halo). The halo of M31 is visible
only for $\sigma_{\rm L}=2$. M31 is located at  $l \approx 122^{\rm o}$ and
$b\approx -21^{\rm o}$.  
on an Aitoff-Hammer projection in Galactic coordinates
}
\end{figure*}

Comparing results for different halo cutoffs, 
we found that all constraints are slightly stronger
for the 300 kpc cutoff.
Here we present the results for the 800 kpc cutoff only,
as it represents the case of an infinite halo and gives more 
conservative constraints.

 The hypothesis that all GRBs originate 
from a Galactic halo is inconsistent
with the data for any parameters. 
However, this fact is only of academic interest 
as observations of GRB afterglows showed that at least a substantial part 
of GRBs has a cosmological origin. 

 The results of the Monte-Carlo simulations for different
parameters: $r_c$, see equation (\ref{spatdistrib}), $P_{100}$, and
$\sigma_{\rm L}$, see equation (\ref{lumin}),
are shown in Figure 2 as maps of isocontours
of the allowed halo fraction $\xi$.
The highest allowed fraction $\xi$ is $0.63$ for $\sigma_{\rm L}=1$
and $0.55$ for $\sigma_{\rm L}=2$ (for $\sigma_{\rm L} = 0$, $\xi = 0.7$,
but the case of $\sigma_{\rm L} = 0$, i.e., for a
standard candle GRB luminosity, is hardly realistic,
and we do not discuss it further).
The brightness intervals where
the fraction of the galactic subpopulation can exceed 0.5 are limited to:
$0.3 <P_{100} < 1.0$ for $\sigma_{\rm L}=1$ and $0.012 < P_{100} < 0.03$
for $\sigma_{\rm L}=2$. In terms of the average peak luminosity
these intervals are
(1.0 - 3.3)$\times 10^{41}$ erg s$^{-1}$ and
(1.8 - 4.5)$\times 10^{40}$ erg s$^{-1}$, respectively
(note the difference between arithmetical mean and logarithmic mean
for the case if lognormal distribution).

 The total rate of GRB events in the halo required to reproduce 
the observed GRB data depends on $\sigma_{\rm L}$.
In the case of $\sigma_{\rm L}=1$, the rate of GRBs
out to 800 kpc should be $\sim$ 8000 per year ($N_0 = 0.004$ kpc$^{-3}$). 
In the case $\sigma_{\rm L}=2$, this number is $\sim$ 45 000 
($N_0 = 0.02$ kpc$^{-3}$).
Nevertheless, the total power emitted in the form of GRBs implied in the two 
cases is almost the same: 
$\sim 2 \cdot 10^{45} T$ erg year$^{-1}$, where $T$
is the average duration of GRBs (a reasonable estimate is $T \sim 10$ s).  
In both cases, M31 is at the edge of the sampling volume. The concentration
of GRBs around the M31 location should be still visible for $\sigma_{\rm L}=2$ 
and hardly visible for $\sigma_{\rm L}=1$ (see Fig. 3).

\section{Constraints on the Euclidean subpopulation}

We used the same logic as in \S3 to constrain the possible
fraction of a GRB subpopulation homogeneously distributed in space
(the Euclidean component).

We represent the whole sample of GRBs as 
$(\xi-1)S_0 + \xi S_e$, where $S_0$ is an isotropic 
subsample (presumably cosmological) with an
unknown log $N$ - log $P$ distribution,
$S_e$ is the homogeneous (in Euclidean space) subsample, 
and $\xi$ is the possible fraction
of GRBs of the homogeneous subsample in the full sample. 
For constraining $\xi$, we used only criterion 4 from \S3.2.

The upper limit on an Euclidean component
is 23\% for the full sample of the UC
(3906 GRBs with peak fluxes down to 0.1 photons cm$^{-2}$ s$^{-1}$).
For the sample from the UC with peak fluxes 
down to the BATSE trigger threshold
 ($\approx$ 0.2 photons cm$^{-2}$ s$^{-1}$),
the limit is about 14\%.     
The same estimate of Kommers et al. (2000) is about 10\%.
They used their log $N$ - log $P$ distribution with fluxes down to about 
0.18-0.20 photons cm$^{-2}$ s$^{-1}$ and
the difference in the constraints on the Euclidean component results from
the difference in the estimates of log $N$ - log $P$ distribution.
Kommers' estimate for the number of dim GRBs is lower.

This constraint on the Euclidean component should
be applied to any kind of local Galactic GRB 
subpopulation ($r$ $<$ 300 pc) and to extragalactic sources at 
small redshifts ($z\ll 1$).
If the sample of GRBs has a 
$\sim$ 20\% fraction of an Euclidean subpopulation, then 
this population dominates 
among weak GRBs. If this subpopulation is intrinsically different, 
it can be revealed in correlations between some GRB properties and 
the brightness.

\section{Conclusions}

The general results of our analysis of the new uniform catalog of GRBs
can be formulated as follows: 

-the large scale isotropy of the GRB sky distribution is confirmed
for the larger and deeper sample of 3906 BATSE GRBs
with peak fluxes down to $\approx$ 0.1 photons cm$^{-2}$ s$^{-1}$;

-the possible fraction of GRBs from an extended galactic halo  
can be up to $\sim$60\%. This is, however, only for narrow intervals 
of the parameters;

-the Euclidean component limited by a fraction of GRBs above
the threshold 0.1 photons cm$^{-2}$ s$^{-1}$ is about 23\%.

These constraints are still weak, so a considerable
fraction of even long GRBs can be prescribed 
to some kind of a non-cosmological source population. 
Stronger constraints can hardly be obtained within
the framework of the statistical approach.
We believe that tighter constraints 
can be imposed from future afterglow observations. 

{\bf Acknowledgement}

This work was  supported by the Swedish 
Natural  Science Research Council, the Royal Swedish Academy of Science,
the Wenner-Gren Foundation for  Scientific  Research,  
and the Russian Foundation for Basic Research (grant 00-02-16135).

\end{document}